\documentclass{IOS-Book-Article}

\usepackage{graphicx}
\usepackage{mathptmx}

\usepackage{xspace}

\def\hb{\hbox to 10.7 cm{}}

\newcommand{\exanest}{ExaNeSt\xspace}

\newcommand{\nvidia}{NVIDIA\xspace}

\newcommand{\ie}{\textit{i.e.}\xspace}
\newcommand{\eg}{\textit{e.g.}\xspace}

\newcommand{\wavescales}{WaveScalES\xspace}
\newcommand{\dpsnn}{DPSNN\xspace}

\newcommand{\minidpsnn}{miniDPSNN\xspace}

\newcommand{\gbe}{GbE\xspace}

\newcommand{\lif}{Leaky Integrate-and-Fire\xspace}
\newcommand{\dpsnnlong}{Distributed and Plastic Spiking Neural Network\xspace}

\begin{document}

\begin{center}
The final publication is available at IOS Press through \\ 
http://dx.doi.org/10.3233/978-1-61499-843-3-760 \\
(2018) Advances in Parallel Computing, 32, pp. 760-769, Talk at ParCo 2017.
\end{center}

\pagestyle{headings}
\def\thepage{}

\begin{frontmatter}              

\title{The brain on low power architectures} 
\subtitle{Efficient simulation of cortical slow waves and asynchronous states}


\author[A]{\fnms{Roberto}        \snm{Ammendola}}
\author[B]{\fnms{Andrea}         \snm{Biagioni}
\thanks{Corresponding Author: Andrea Biagioni, INFN Sezione di Roma, Piazzale Aldo Moro 2, Roma, Italy\\  
E-mail:andrea.biagioni@roma1.infn.it}},
\author[B]{\fnms{Fabrizio}       \snm{Capuani}}
\author[B]{\fnms{Paolo}          \snm{Cretaro}}
\author[B]{\fnms{Giulia}         \snm{De Bonis}}
\author[B]{\fnms{Francesca}      \snm{Lo Cicero}}
\author[B]{\fnms{Alessandro}     \snm{Lonardo}}
\author[B]{\fnms{Michele}        \snm{Martinelli}}
\author[B]{\fnms{Pier Stanislao} \snm{Paolucci}}
\author[B]{\fnms{Elena}          \snm{Pastorelli}}
\author[B]{\fnms{Luca}           \snm{Pontisso}}
\author[B]{\fnms{Francesco}      \snm{Simula}}
and
\author[B]{\fnms{Piero}          \snm{Vicini}}

\address[A]{INFN, Sezione di Roma Tor Vergata, Italy}
\address[B]{INFN, Sezione di Roma, Italy}

\begin{abstract}
Efficient brain simulation is a scientific grand challenge, a
parallel/distributed coding challenge and a source of requirements and
suggestions for future computing architectures.
Indeed, the human brain includes about $10^{15}$ synapses and
$10^{11}$ neurons activated at a mean rate of several Hz.
Full brain simulation poses Exascale challenges even if simulated at
the highest abstraction level.
The \wavescales experiment in the Human Brain Project (HBP) has the
goal of matching experimental measures and simulations of slow waves
during \mbox{deep-sleep} and anesthesia and the transition to other
brain states.
The focus is the development of dedicated \mbox{large-scale}
parallel/distributed simulation technologies.
The \exanest project designs an \mbox{ARM-based}, \mbox{low-power} HPC
architecture scalable to million of cores, developing a dedicated
scalable interconnect system, and SWA/AW simulations are included
among the driving benchmarks.
At the joint between both projects is the INFN proprietary Distributed
and Plastic Spiking Neural Networks (\dpsnn) simulation engine.
DPSNN can be configured to stress either the networking or the
computation features available on the execution platforms.
The simulation stresses the networking component when the neural net
--- composed by a relatively low number of neurons, each one
projecting thousands of synapses --- is distributed over a large
number of hardware cores.
When growing the number of neurons per core, the computation starts to
be the dominating component for short range connections.
This paper reports about preliminary performance results obtained on
an \mbox{ARM-based} HPC prototype developed in the framework of the
\exanest project.
Furthermore, a comparison is given of instantaneous power, total
energy consumption, execution time and energetic cost per synaptic
event of SWA/AW DPSNN simulations when executed on either ARM- or
\mbox{Intel-based} server platforms.
\end{abstract}

\end{frontmatter}
\markboth{October 2017\hb}{October 2017\hb}

\section{Introduction}

The scaling of the performance of modern HPC systems and applications
is strongly limited by the energy consumption.
Electricity is the main contributor to the total cost of running an
application and \mbox{energy-efficiency} is becoming the principal
requirement for this class of computing devices.
In this context, the performance assessment of processors with a high
\mbox{performance-per-watt} ratio is necessary to understand how to
make \mbox{energy-efficient} computing systems for scientific
applications. 
Processors based on the ARM architecture dominate the market of
\mbox{low-power} and \mbox{battery-powered} devices such as tablets
and smartphones.
Several scientific communities are exploring \mbox{non-traditional}
\mbox{many-core} processors architectures coming from the embedded
market, from the Graphics Processing Unit (GPU) to the
\mbox{System-on-Chip} (SoC), looking for a better tradeoff between
\mbox{time-to-solution} and \mbox{energy-to-solution}.
A number of research projects are active in trying to design an actual
platform along this direction.
The \mbox{Mont-Blanc} project
~\cite{montblanc:2016:short,montblanc:2017:Online}, coordinated by the
Barcelona Supercomputing Center, has deployed two generations of HPC
clusters based on ARM processors, developing also the corresponding
ecosystem of HPC tools targeted to this architecture.
Another example is the \mbox{EU-FP7}
EUROSERVER~\cite{Marazakis:EUROSERVER:2016} project, coordinated by
CEA, which aims to design and prototype technology, architecture, and
systems software for the next generation of datacenter
``microservers'', exploiting \mbox{64-bit} ARM cores.

Fast simulation of spiking neural network models plays a dual role:
(i) it contributes to the solution of a scientific grand challenge ---
\ie the comprehension of brain activity --- and, (ii) by including it
into embedded systems, it can enhance applications like autonomous
navigation, surveillance and robotics.
Therefore, these simulations assume a driving role in shaping the
architecture of either specialized and \mbox{general-purpose}
\mbox{multi-core}/\mbox{many-core} systems to come, standing at the
crossroads between embedded and High Performance Computing.
See, for example~\cite{Merolla668:short}, describing the TrueNorth
\mbox{low-power} specialized hardware architecture dedicated to
embedded applications, and~\cite{Stromatias:2013} discussing the power
consumption of the SpiNNaker hardware architecture, based on embedded
\mbox{multi-cores}, dedicated to brain simulation.
Worthy of mention are also~\cite{gewaltig:2007,Modha:2011} as examples of
approaches based on standard HPC platforms and \mbox{general-purpose}
simulators.

The APE Research Group at INFN developed a distributed neural network
simulator~\cite{Paolucci:2013:Distributed} as a
\mbox{mini-application} and benchmark in the framework of the EURETILE
FP7 project~\cite{EURETILE:JSA:2016:short}.
Indeed, the \dpsnnlong with synaptic \mbox{Spike-Timing} Dependent
Plasticity \mbox{mini-application} was developed with two main
purposes in mind: as a quantitative benchmarking tool for the
evaluation of requirements for future embedded and HPC systems and as
an efficient simulation tool addressing specific scientific problems
in computational neuroscience.
As regards the former goal, the \exanest
project~\cite{DSD:EXANEST:2016:short} includes \dpsnn in the set of
benchmarks used to specify and validate the requirements of future
interconnects and storage systems;
as an example of the latter, the distributed simulation technology is
employed in the study of slow waves in large scale cortical
fields~\cite{ruiz:2011,stroh:2013} in the framework of HBP project.

This paper describes porting \dpsnn onto different \mbox{ARM-based}
platforms and running it on \mbox{low-power} CPUs, comparing the
resulting computing and energy performances with traditional systems
mainly based on x86 multicores.
The characterization of \mbox{\dpsnn-generated} data traffic is
described, highlighting the limitations faced when the application is
run on \mbox{off-the-shelf} networking components.
The code organization and its compactness give the \dpsnn a high
degree of tunability, giving the opportunity to test different areas
of the platform.
The networking compartment is the most stressed when the simulated
neural net --- composed by a relatively low number of neurons, each
one projecting thousands of synapses --- is distributed over a large
number of hardware cores.
When the number of neurons per core grows, the impact of both computing
and memory increases.
For this reason, we employ \dpsnn as a general benchmarking tool for
HPC systems.

\section{\mbox{Mini-application} benchmarking tool}

Evaluation of HPC hardware is a key element especially in the first
stages of a project --- \ie definition of specification and design ---
and during the development and implementation.
Features impacting performance should be identified in the analysis
and design of new architectures.
In the early stages of the development, full applications are too
complex to run on the hardware prototype.
In usual practice, hardware is tested with very simple kernels and
benchmarking tools which often reveal their inadequacy as soon as they
are compared with real applications running on the final platform,
showing a huge performance gap.

In the last years, a new category of compact, \mbox{self-contained}
proxies for real applications called \textit{mini-apps} has appeared.
Although a full application is usually composed by a huge amount of
code, the overall behaviour is driven by a relatively small subset.
\mbox{Mini-apps} are composed by these core operations providing a
tool to study different subjects:
(i) analysis of the computing device --- \ie the node of the system.
(ii) evaluation of scaling capabilities, configuring the
\mbox{mini-apps} to run on different numbers of nodes, and
(iii) study of the memory usage and the effective throughput towards
the memory.

This effort is led by the Mantevo project~\cite{heroux:2009}, that
provides application performance proxies since 2009.
Furthermore, the main research computing centers provide sets of
\mbox{mini-applications}, adopted when procuring the systems, as in
the case of
the \mbox{NERSC-8}/Trinity Benchmarks~\cite{Cordery:2014}, used to
assess the performance of the Cray XC30 architecture, or
the Fiber Miniapp Suite~\cite{fiber}, developed by RIKEN Advanced
Institute for Computational Science (RIKEN AICS) and the Tokyo
Institute of Technology.

The \minidpsnn benchmarking tool leverages on the
\mbox{Hardware-Software} \mbox{Co-design} approach that starts from
the collection of application requirements for the initial development
of the infrastructure and then pursues testing the adopted solution
during the implementation.
Thus, the application drives the research about the main components of
a HPC system from its roots by optimizing modeling and simulation of a
complex system.

The analysis is based on the behaviour of a strong scaling test.
Neurons are arranged into ``columns'', each one composed by about one
thousand neurons; columns are then arranged into a bidimensional grid.
Each excitatory neuron projects 80\% of its synapses out to those
residing in its own column while the rest reaches out to those in the
neighbouring columns, according to the chosen remote connectivity.
Instead, synapses of inhibitory neurons are projected only towards
excitatory ones residing in their same column.
When \dpsnn runs, each process can either host a fraction of a column,
a whole single column, or an integer number of columns.

Each core of the computing system hosts only one process optimizing
the performance.
Thus, the varying of the \mbox{columns-per-process} ratio --- \ie
ratio of columns per core of the computing devices --- throttles the
application into different regimes, allowing to stress and test
several elements of the platform.
Be noted that  in general, the hardware connection topology bears no
resemblance whatsoever with the lateral connectivity of columns and
neurons, the exception being when running only one process per node,
so that all outwards connectivity of a column impinges upon the
network system of the node.


Here is a rundown of the application tasks that \minidpsnn performs
and that allow to gauge the components of the architecture under test:
\begin{itemize}
\item \textbf{Computation}: processing of the time step in the
  dynamical evolution of the neuron.
\item \textbf{Memory Management}: management of either axonal spikes
  organized in time delay queues and lists of synaptic spikes, both
  stored in memory.
\item \textbf{Communication}: transmission along the interconnect
  system of the axonal spikes to the subset of processes where target
  neurons exist.
\item \textbf{Synchronization}: at each time step, the processes
  deliver the spikes produced by the dynamics according to the
  internal connectivity supported by the synaptic configuration. This
  global exchange is currently implemented by means of synchronous MPI
  collectives; any offset in time when different processes reach these
  waypoints --- whether it be by fluctuations in load or network
  congestion --- causes idling cores and diminished parallelization.
\end{itemize}

\begin{table}[!hbt]
\footnotesize
\centering
\begin{tabular}{|l|c|c|c|}
\hline
\textbf{Grid}               & \textbf{$12 \times 12$} & \textbf{$24 \times 24$} & \textbf{$48 \times 48$} \\ 
\hline
\textbf{Neurons}            & 0.18~M                  & 0.71~M                  & 2.86~M        \\                     
\hline
\textbf{Synapses}           & 0.20~G                  & 0.80~G                  & 3.20~G        \\                     
\hline
\textbf{Columns}            & 144                     & 192                     & 192           \\                     
\hline
\textbf{Columns/Core}       & \textbf{1}              & \textbf{3}              & \textbf{12}   \\                     
\hline
\textbf{Simulated Seconds}  & 30                      & 12                      & 18            \\                     
\hline
\textbf{Wall-clock Seconds} & 1484                    & 2148                    & 15182         \\                     
\hline
\textbf{Computation}        & 21.3\%                  & 34.2\%                  & 45.1\%        \\                     
\hline
\textbf{Memory Management}  & 17.1\%                  & 16.7\%                  & 16.9\%        \\                     
\hline
\textbf{Communication}      & 35.2\%                  & 10.7\%                  & 0.9\%         \\                     
\hline
\textbf{Synchronization}    & 22.9\%                  & 36.3\%                  & 36.2\%        \\                     
\hline
\end{tabular}
\caption{\minidpsnn tasks overview.}
\label{tab:dpsnntask}
\end{table}

Table~\ref{tab:dpsnntask} displays results obtained running on a
standard HPC cluster based on Intel Xeon processors communicating over
an InfiniBand interconnect, as a function of the configuration of the
testbed --- \ie grid size, simulated seconds, allocated cores.
The distribution of tasks is strongly dependent on the
\mbox{columns-per-core} ratio.
As already stated, the computation task becomes more demanding when
increasing the number of columns per node --- which means increasing
the total number of neurons.
Instead, reducing the \mbox{columns-per-core} ratio generates
relatively more communication among processes, moving the focus of the
test to the interconnect.

\subsection{Analysis of \mbox{low-power} and off-the-shelf architectures in the \mbox{real-time} domain}

In this domain, being ``real-time'' signifies a \minidpsnn workpoint
such that the execution time --- \ie \mbox{wall-clock} time of the
running application --- is not greater than the simulated time.
Accomplishment of this workpoint is obtained through an accurate
configuration of parameters.
Prelimary trials of \dpsnn keeping pace with this \mbox{real-time}
requirement are reported in this section.
This working condition could be useful in the robotics application
field.

\begin{figure}[!hbt]
\centering
  \begin{minipage}[t]{.48\textwidth}
    \centering    
    \includegraphics[width=.95\textwidth]{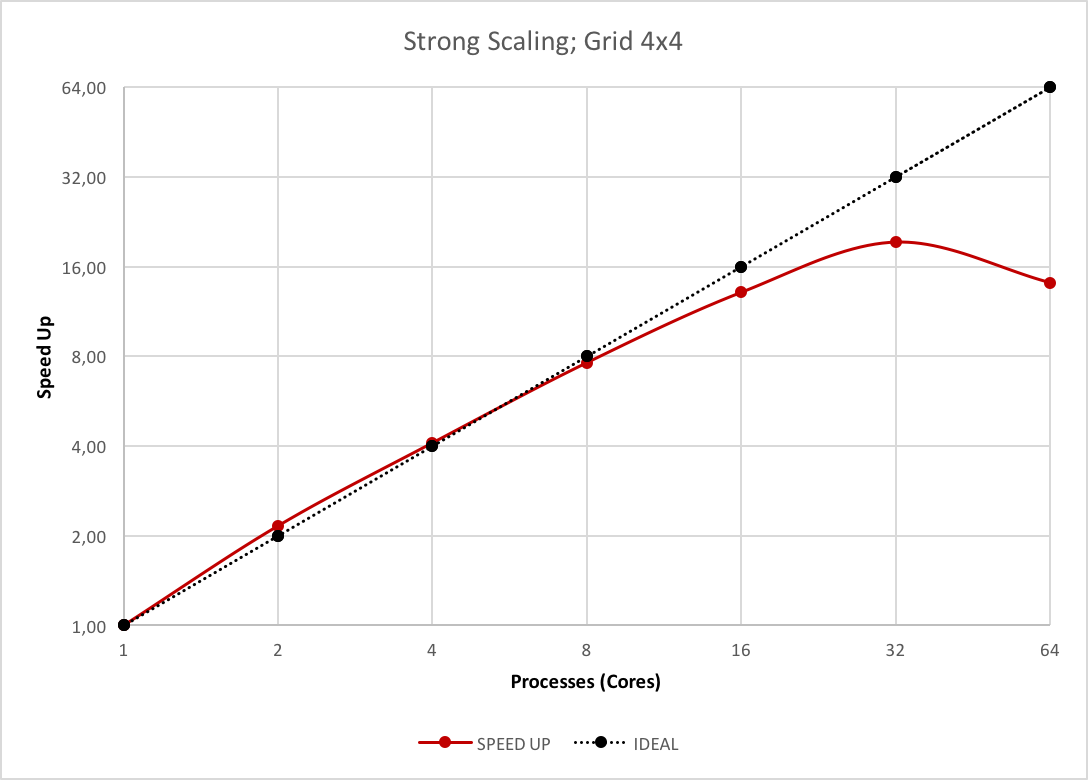}
    \caption{Strong scaling of a $4\times4$ column grid simulated on
      an \mbox{Intel-based} platform equipped with IB.}
    \label{fig:dpsnn4x4}
  \end{minipage}
  \quad
  \begin{minipage}[t]{.48\textwidth}
    \centering    
    \includegraphics[width=.95\textwidth]{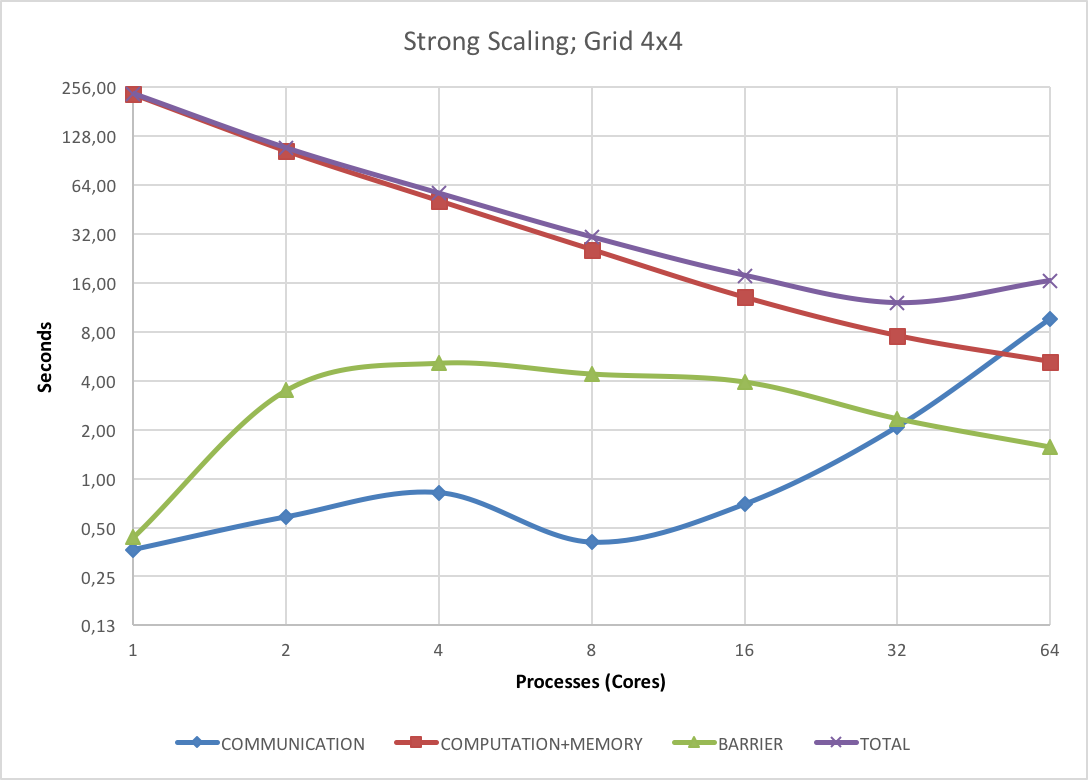}
    \caption{\minidpsnn analysis of the \mbox{Intel-based} platform.}
    \label{fig:dpsnn4x4task}
  \end{minipage}	
\end{figure}

The testbed is a standard strong scaling test of a $4\times4$ columns
grid.
Figure~\ref{fig:dpsnn4x4} shows the results of the test obtained
simulating 10~s on the \mbox{Intel-based} platform.

Up to $8\div16$ cores, the architecture scales well, decreasing the
execution time down to $\sim12$~seconds.
The execution time increases unexpectedly ($\sim16$~seconds) when
distributing the problem over 32 cores, thus preventing the
achievement of the target workpoint.

Singling out the times of the various tasks as reported in
Figure~\ref{fig:dpsnn4x4task} sheds some light on this behaviour.
We see that communication quickly becomes more demanding when the
problem is split over more than 16 processes, dominating the behaviour
of the application.
As mentioned before, the application stresses the interconnect when
the \mbox{column-per-core} ratio decreases --- a whole column or
portions of column are managed by each core in the tested
configuration.
More than 80\% of synapses remain within the column their projecting
neuron belongs to.
Communication between processes increases when the columns are split
among them, clogging the network with an ever increasing number of
small packets.
The \minidpsnn highligths this ``latency'' limitation of the IB
interconnect provided by the cluster.
In general, COTS interconnects offer adequate throughput when moving
large amounts of data, but tipically trudge when the communication is
\mbox{latency-dominated}.
This issue with communication --- manifesting here with a number of
computing cores which is, by today's standards, not large --- is
similar to that encountered by the parallel cortical simulator
C2~\cite{Ananthanarayanan:2009} --- targeting a scale in excess of
that of the cat cortex --- on the Dawn Blue Gene/P supercomputer at
LLNL, with 147456 CPUs and 144 TB of main memory.
The capability to replicate the behaviour of a supercomputer with a
\mbox{mini-app} running on a limited number of 1U servers could be
considered the proof of its effectiveness.

\begin{figure}[!hbt]
\centering
  \begin{minipage}[t]{.48\textwidth}
    \centering    
    \includegraphics[width=.95\textwidth]{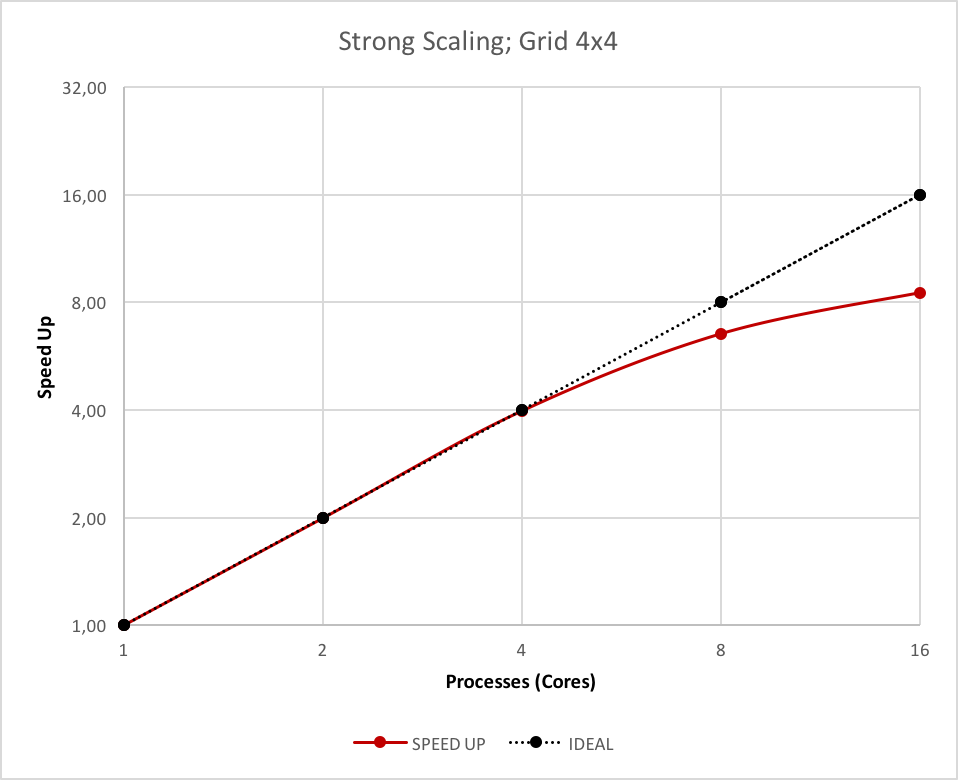}
    \caption{Strong scaling of a $4\times4$ column grid simulated on
      an \mbox{ARM-based} platform equipped with \gbe interconnect.}
    \label{fig:dpsnn4x4trenz}
  \end{minipage}
  \quad
  \begin{minipage}[t]{.48\textwidth}
    \centering    
    \includegraphics[width=.95\textwidth]{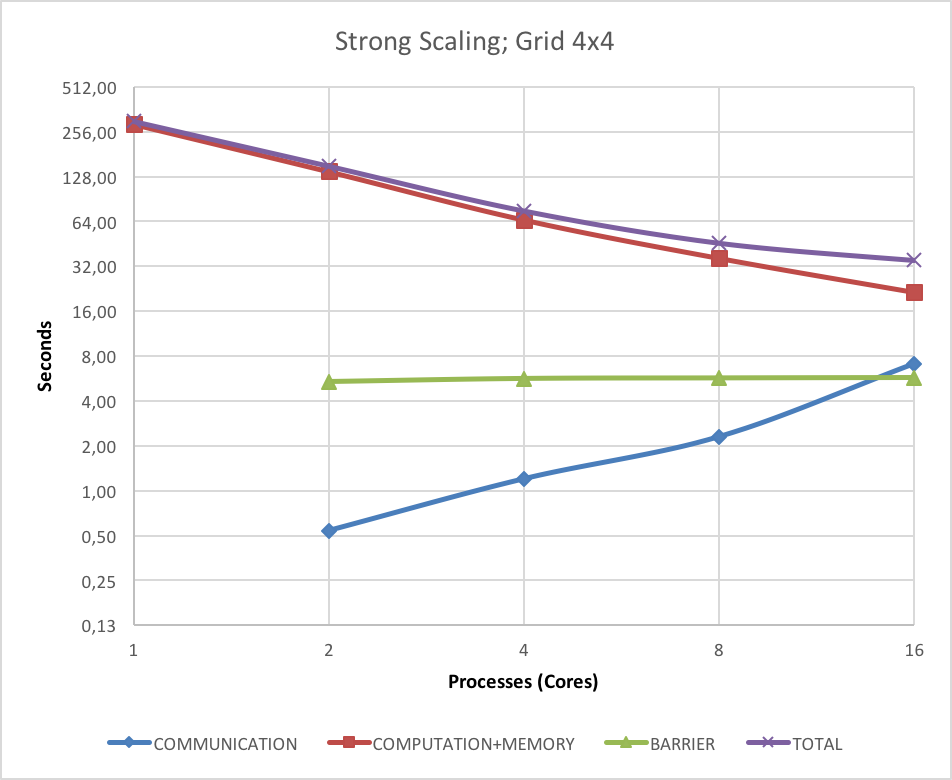}
    \caption{\minidpsnn analysis of the \mbox{ARM-based} platform.}
    \label{fig:dpsnn4x4trenztask}
  \end{minipage}	
\end{figure}

Similar results are obtained performing the same test on an
\mbox{ARM-based} platform as showed in Figure~\ref{fig:dpsnn4x4trenz}
and Figure~\ref{fig:dpsnn4x4trenztask}, although the analysis is
limited by the available number of cores (16).
The \mbox{ARM-based} prototype is composed by four nodes, each node
consisting of a TEBF0808 Trenz board equipped with a Trenz TE0808
UltraSOM+ module.
The Trenz UltraSOM+ consists of a Xilinx Zynq UltraScale+
\mbox{xczu9eg-ffvc900-1-e-es1} MPSoC and 2~Gbytes of DDR4 memory.
The Zynq UltraScale+ MPSoC incorporates both a processing system
composed by \mbox{quad-core} ARM \mbox{Cortex-A53} and the
programmable logic --- not used in this test.
All four nodes are connected together through a 1~Gbps
\mbox{Ethernet-based} network.

\begin{figure}[!hbt]
\centering
  \begin{minipage}[t]{.48\textwidth}
    \centering    
    \includegraphics[width=.95\textwidth]{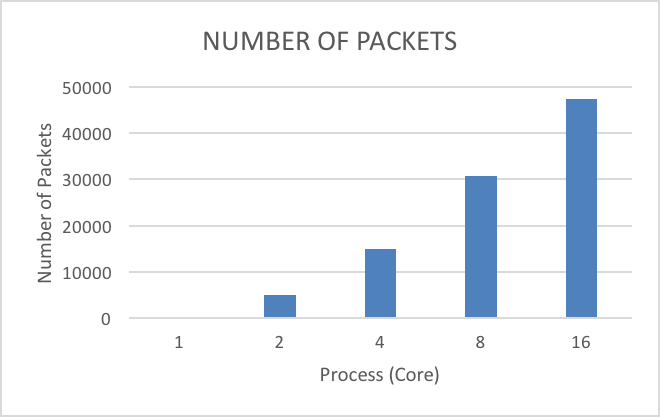}
    \caption{Packets generated during the simulation.}
    \label{fig:dpsnnpackets}
  \end{minipage}
  \quad
  \begin{minipage}[t]{.48\textwidth}
    \centering    
    \includegraphics[width=.95\textwidth]{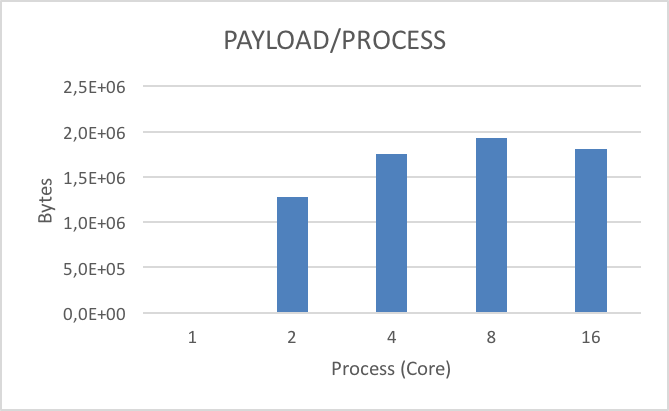}
    \caption{Payload generated by each process.}
    \label{fig:dpsnnpayload}
  \end{minipage}
\end{figure}

The number of transmitted packets increases distributing the same
problem over an increasing number of processes (cores) as shown in
Figure~\ref{fig:dpsnnpackets}
while the payload generated by each process does not vary as shown in
Figure~\ref{fig:dpsnnpayload} and the communication becomes more
demanding.

\begin{figure}[!hbt]
\centering
  \begin{minipage}[t]{.48\textwidth}
    \centering    
    \includegraphics[width=.95\textwidth]{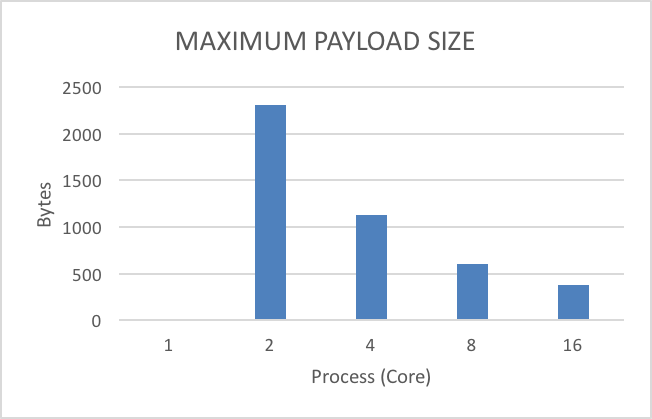}
    \caption{Maximum packet size produced the \dpsnn
	simulation.}
    \label{fig:dpsnnmax}
  \end{minipage}
  \quad
  \begin{minipage}[t]{.48\textwidth}
    \centering    
    \includegraphics[width=.95\textwidth]{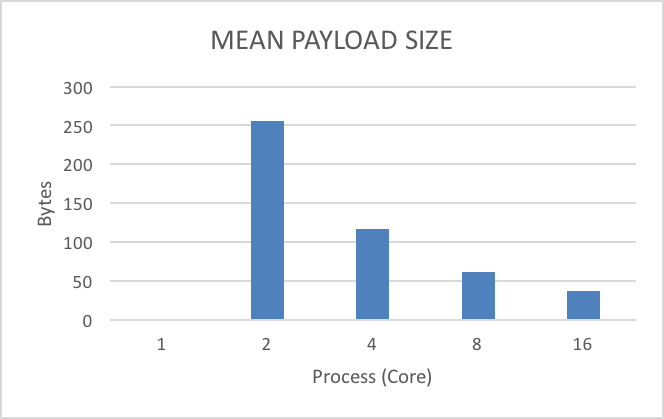}
    \caption{Mean packet size produced the \dpsnn
	simulation.}
    \label{fig:dpsnnmean}
  \end{minipage}	
\end{figure}

Furthermore, the size of packets decreases (see
Figure~\ref{fig:dpsnnmax}); the mean packet size is $\sim40$~bytes
when each core simulates the dynamics of a single column, as depicted
in Figure~\ref{fig:dpsnnmean}.

The characterization of the traffic generated by the \dpsnn over
several \mbox{off-the-shelf} interconnects allows to identify the main
requirement for a network device of future exascale computing system
simulating spiking neural network simulation: the network system
should be optimized for the trasmission of small packets.
In particular, performances are strongly influenced by
(i) the design and implementation of a \mbox{low-latency} interconnect
architecture, and
(ii) the definition of a light and reliable communication protocol
guaranteeing high throughput and optimizing the transfers of data
packets with payload $<512$~Bytes.

Finally, a planned \mbox{re-engineering} of the \dpsnn foresees a
\mbox{two-level} hierarchy enforced via MPI communicators:
one auxiliary process (called ``broker'') is added per node and
communications are segregated to be only among processes belonging to
the same node –-- \ie exchanges that go only through
\mbox{intra-node}, \mbox{shared-memory} channels –-- or among brokers
–-- \ie exchanges that only go through \mbox{inter-node}, remote
interfaces.
In this way, ``local'' exchanges among neighbouring neural columns
(which, given the biologically plausible topology for the synaptic
connectivity, make up the exchange bulk) can be contained to the
fastest and possibly less congested \mbox{intra-node} channel while
``distal'' exchanges are gathered to the broker process of the node,
then scattered to brokers of other nodes that take care of scattering
them to the appropriate recipients.

\section{\mbox{Energy-to-Solution} analysis}

Instantaneous power, total energy consumption, execution time and
energetic cost per synaptic event of a spiking neural network
simulator distributed on MPI processes are compared when executed on
different generations of \mbox{low-power} and traditional computing
architecture to have a (limited) estimate of the trend.

The power and energy consumption reported were obtained simulating 3~s
of activity of a network made of 18~M equivalent (internal + external)
synapses:
the network includes 10~K neurons (\lif with \mbox{Calcium-mediated}
\mbox{spike-frequency} adaptation), each one projecting an average of
1195 internal synapses and receiving an ``external'' stimulus,
corresponding to 594 equivalent external \mbox{synapses/neuron}.
A Poissonian spike train targets external synapses with an average
rate of 3~Hz; synaptic plasticity is disabled.
In response, the neurons fire trains of spikes at a mean rate of
5.1~Hz.

The power measurement equipment consists of a DC power supply, a
\mbox{high-precision} Tektronix DMM4050 digital multimeter for DC
current measurements connected to National Instruments data logging
software and a \mbox{high-precision} AC power meter.
The AC power of the \mbox{high-end} server node is measured by a
Voltech PM300 Power Analyzer upstream of the main server power supply
(measuring on the AC cable).
For the SoCs, the DC current was instead sampled downstream of the
power supply.
Such difference should not affect significantly the results, given the
closeness to one of the $\cos \varphi$ factor of the server power
supply.

\subsection{First Generation Benchmark}

The traditional computing system --- \ie ``server platform'' --- is
based on a SuperMicro \mbox{X8DTG-D} 1U \mbox{dual-socket} server
housing 8 computing cores residing on \mbox{quad-core} Intel Xeon CPUs
(Westmere E5620@2.4~GHz in 32~nm CMOS technology).
This ``server platform'' is juxtaposed to the ``embedded platform'':
two \nvidia Jetson TK1 boards, connected by an Ethernet 100~Mb
\mbox{mini-switch} to emulate a \mbox{dual-socket} node, each board
equipped with a \nvidia Tegra K1 chip, \ie a \mbox{quad-core} ARM
\mbox{Cortex-A15}@2.3~GHz in 28~nm CMOS technology.

The ``server platform'' has 48~GB of DDR3 memory \mbox{on-board},
operating at 1333~MHz --- 6~GB per core --- while the ``embedded
platform'' only has 2~GB running at 933~MHz --- 0.5~GB per core.
This makes for a considerable difference in terms of memory bandwidth
--- 14.9~GB/s for the \mbox{ARM-based} system against the 25.6~GB/s of
the \mbox{Intel-based} one --- which has an impact on \dpsnn and its
intensive memory usage, \eg for delivering spikes to
\mbox{post-synaptic} neuron queues.

\begin{figure}[!hbt]
\centering
  \begin{minipage}[t]{.47\textwidth}
    \centering
    \includegraphics[width=.95\textwidth]{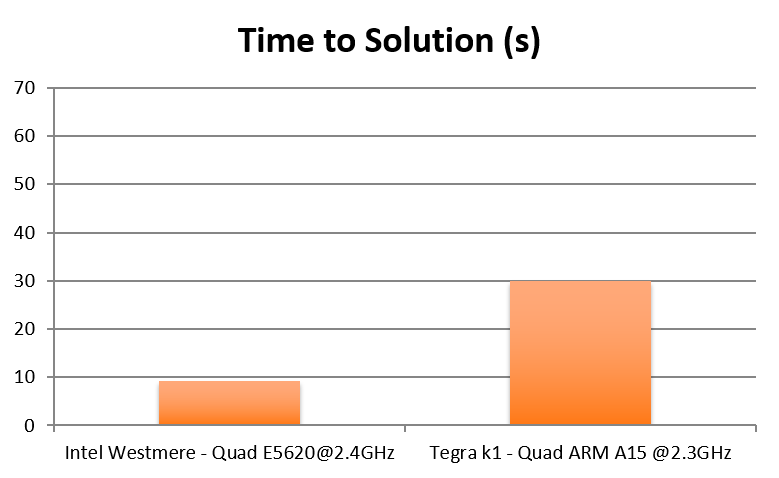}
    \caption{First generation \mbox{time-to-solution} result.}
    \label{fig:dpsnntime_first}
  \end{minipage}
  \quad
  \begin{minipage}[t]{.47\textwidth}
    \centering
    \includegraphics[width=.95\textwidth]{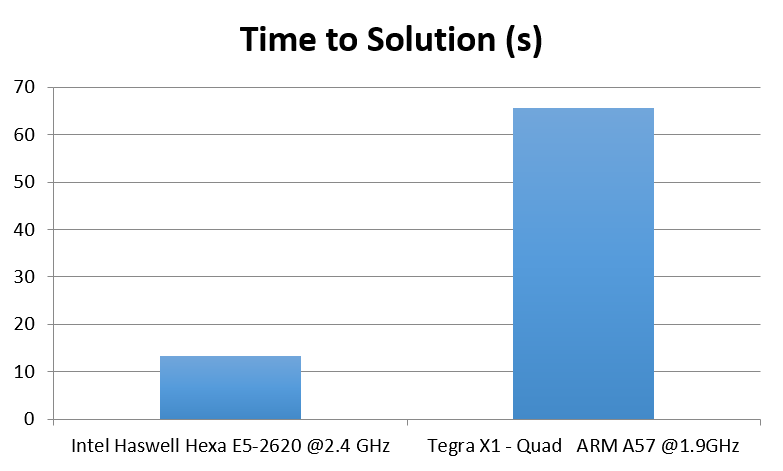}
    \caption{Second generation \mbox{time-to-solution} result. Note 
	that the number of cores used in the first generation was the 
	double of that used in this case.}
    \label{fig:dpsnntime_second}
  \end{minipage}  
\end{figure}

Partitioning the neural grid onto 8~MPI processes, the simulation of
3~s of activity required 9.1~s on the ``server platform'' and 30~s on
the ``embedded platform'', as shown in
Figure~\ref{fig:dpsnntime_first}.

\begin{figure}[!hbt]
\centering
  \begin{minipage}[t]{.47\textwidth}
    \centering
    \includegraphics[width=.95\textwidth]{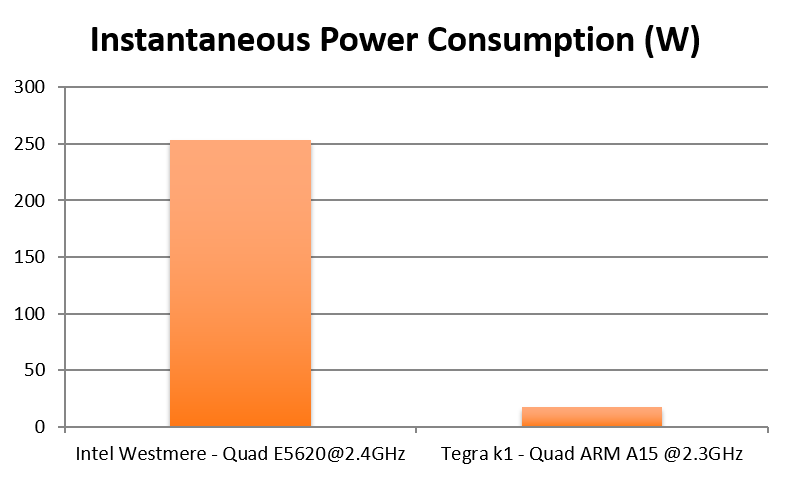}
    \caption{First generation \mbox{power-to-solution} result.}
    \label{fig:dpsnnpower_first}
  \end{minipage}
  \quad
  \begin{minipage}[t]{.47\textwidth}
    \centering
    \includegraphics[width=.95\textwidth]{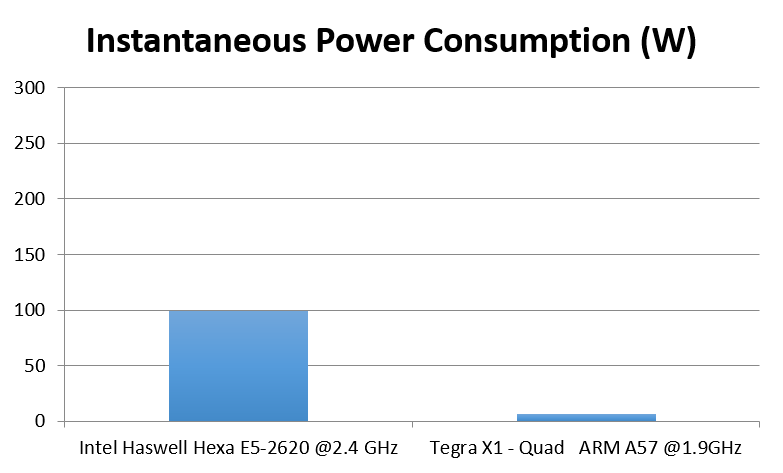}
    \caption{Second generation \mbox{power-to-solution} result. Note 
	that the number of cores used in the first generation was the 
	double of that used in this case.}
    \label{fig:dpsnnpower_second}
  \end{minipage}  
\end{figure}

\begin{figure}[!hbt]
\centering
  \begin{minipage}[t]{.47\textwidth}
    \centering
    \includegraphics[width=.95\textwidth]{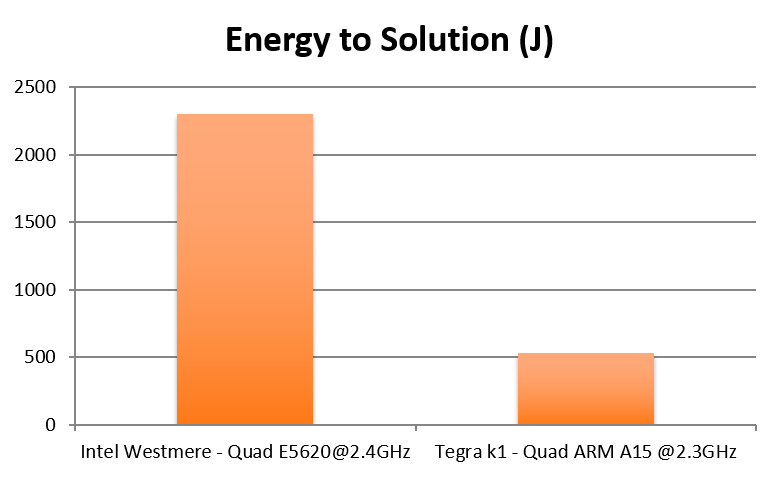}
    \caption{First generation \mbox{energy-to-solution}.}
    \label{fig:dpsnnenergy_first}
  \end{minipage}
  \quad
  \begin{minipage}[t]{.47\textwidth}
    \centering
    \includegraphics[width=.95\textwidth]{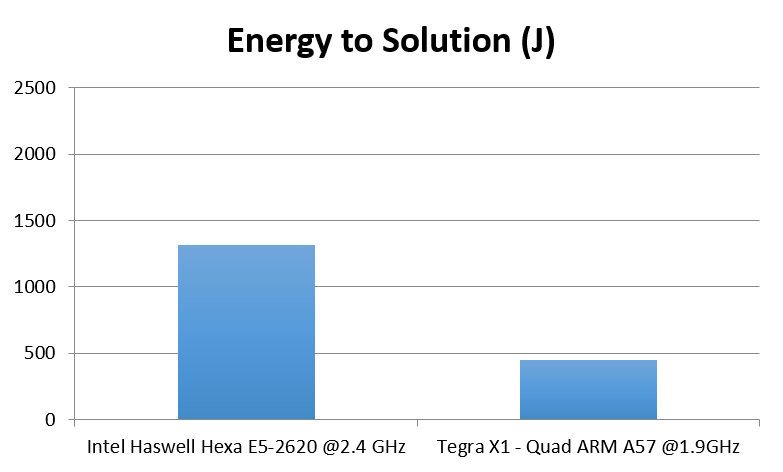}
    \caption{Second generation \mbox{energy-to-solution}.}
    \label{fig:dpsnnenergy_second}
  \end{minipage}  
\end{figure}

Observed currents were $I_{s} = 1.15$~A (``server'') and $I_{e} =
80$~mA (``embedded''), with a 5~mA measure error.
Therefore, the energies required to complete the same task on the two
architectures were $E_{s} = 2.3$~KJ and $E_{e}= 528$~J (see
Figure~\ref{fig:dpsnnenergy_first}), while the observed instantaneous
power consumptions were $P_{s} = 253$~W and $P_{e} = 17.6$~W (see
Figure~\ref{fig:dpsnnpower_first}).
Note that we did not subtract any ``base-line'' power --- \eg power
consumption after bootstrap, so the estimate is ``pessimistic'' in the
sense that it includes the load of the complete system runnning.

The simulation produced a total of 235~M synaptic events: the total
energetic cost of simulation can be estimated in 2.2~$\mu$J/synaptic
event on the ``embedded platform'' node and 9.8~$\mu$J/synaptic
event for the ``server platform''.
The ``server platform'' \mbox{dual-socket} node is faster, spending
3.3 times less time than the ``embedded platform'' node.
However, the ``embedded platform'' node consumes a total energy 4.4
times lower to complete the simulation task, with an instantaneous
power consumption 14.4 times lower than the ``server platform'' node.

The energetic cost of the optimized Compass simulator of the TrueNorth
\mbox{ASIC-based} platform, run on an Intel Core i7 CPU 950@3.07~GHz
(45~nm CMOS process) with 4 cores and 8 threads, is
5.7~$\mu$J/synaptic event, but excludes a significant \mbox{base-line}
power consumption.
Our measures show that if we excluded a similar \mbox{base-line} our
power consumption would be approximately reduced by a factor 4 on 
the ``server platform'', and by a factor 2 on the ``embedded platform''
platform.

\subsection{Second Generation Comparison}

The performances are measured executing the \dpsnn code along with
those of a coeval mainstream Intel processor architecture using a
\mbox{hardware/software} configuration suitable to extrapolate a
direct comparison of \mbox{time-to-solution} and
\mbox{energy-to-solution} at the level of the single core.
The measures are extended to the new generation NVIDIA Jetson TX1 SoC
based on the ARMv8 architecture.
The Jetson TX1 includes four ARM \mbox{Cortex-A57} cores plus four ARM
\mbox{Cortex-A53} cores in big.LITTLE configuration.

The ``server platform'' is a Supermicro SuperServer \mbox{7048GR-TR}
with two \mbox{hexa-core} Intel Haswell \mbox{E5-2620} v3 @2.40~GHz.
Four MPI processes are run on either platform, simulating 3~s of the
dynamics of a network made of $10^{4}$ \lif with Calcium Adaptation
(LIFCA) neurons connected via $18 \times 10^6$ synapses.
Results are shown in Figure~\ref{fig:dpsnntime_second},
Figure~\ref{fig:dpsnnpower_second} and
Figure~\ref{fig:dpsnnenergy_second}.
Although the x86 architecture is about $5 \times$ faster than the ARM
\mbox{Cortex-A57} core in executing the simulation, the energy it
consumes in doing so is $\sim 3 \times$ higher~\cite{cesini:2017}.

\section{Conclusion}

The characterization of the network traffic generated by a cortical
simulator running on a standard computing system provides information
for the definition of specification of custom interconnect.
The result obtained with \minidpsnn drives to the specification of 
a network IP, characterized by a low-latency transfer optimized 
architecture and to the definition of a data transmission protocol
providing high-throughput also for small dimensions of data payload.
Finally ARM processors turned out to be an efficient solution in terms of
power consumption.
The energy-to-solution result obtained running the \dpsnn application
on ARM Cortex-A57 based platform is about three times lower than the
x86 core architecture.

\section*{Acknowledgment}

This work has received funding from the European Union’s Horizon 
2020 Research and Innovation Programme under Grant Agreement No. 720270 
(HBP SGA1) and under Grant Agreement No. 671553 (ExaNeSt).


\bibliographystyle{iopart-num}
\bibliography{../../../ape_bib/bibliography}

\end{document}